\documentclass[epj]{svjour}
\usepackage{graphicx}
\usepackage{amsmath}
\usepackage{amssymb}
\begin{document}

\title{Photonic Properties of Metallic-Mean Quasiperiodic Chains}

\author{Stefanie Thiem\thanks{\email{stefanie.thiem@physik.tu-chemnitz.de}} and Michael Schreiber}
\institute{Institut f\"ur Physik, Technische Universit\"at Chemnitz, D-09107 Chemnitz, Germany}

\abstract{
The light propagation through a stack of two media with different refractive indices, which are aligned according to different quasiperiodic sequences determined by metallic means, is studied using the transfer matrix method. The focus lies on the investigation of the influence of the underlying quasiperiodic sequence as well as the dependence of the transmission on the frequency, the incidence angle of the light wave and different ratios of the refractive indices. In contrast to a periodically aligned stack we find complete transmission for the quasiperiodic systems for a wide range of different refraction indices for small incidence angles. Additional bands of moderate transmission occur for frequencies in the range of the photonic band gaps of the periodic system. Further, for fixed indices of refraction we find a range of almost perfect transmission for angles close to the angle of total reflection, which is caused by the bending of photonic transmission bands towards higher frequencies for increasing incidence angles. Comparing with the results of a periodic stack the quasiperiodicity seems to have only an influence in the region around the midgap frequency of a periodic stack.}

\maketitle

\section{Introduction}\label{sec:introduction}

Since the discovery of quasicrystals by Shechtman et al. in 1984 \cite{PhysRevLett.1984.Shechtman}, many efforts have been conducted to understand the physical properties of these aperiodic materials, which possess a long range order without having a translational symmetry. Hence, quasicrystals are regarded to have a degree of order intermediate between crystals and disordered systems. Quasicrystalline systems have been extensively studied e.g. with respect to their structure, which shows uncommon rotational symmetries \cite{PhysRevLett.1984.Shechtman,PhysRevLett.1987.Wang,PhysRevLett.1985.Ishimasa}, and their electronic states, which were found to show a Cantor-set spectrum in one dimension \cite{JPhysFrance.1989.Sire,PhysRevB.1987.Kohmoto,JPhysA.1989.Gumbs,JStatPhys.1989.Suto}; but also phonons and magnetic properties of these materials have been investigated \cite{ModPhys.1994.Baake,PhysRevLett.2003.Vedmedenko,ZPhysB.1987.Chen,PhysRevB.1986.Nori,Ferro.2004.Ilan}.

Further, the photonic properties are of special interest because the complex symmetries in quasicrystals make them suitable for the application in several optical devices such as single-mode light-emitting diodes, polarization switching and microelectronic devices that are based on photons rather than on electrons, which potentially can be the electromagnetic analogue to semiconductors \cite{PhysWorld.2004.McGrath,JPhysD.2007.Steurer,PhilMag.2008.Bahabad,PhysRevB.2001.Macia,OptExp.2008.Hendrickson}.

For the theoretical study of quasiperiodic systems one often applies the concept of aperiodic mathematical sequences / tilings. Especially, the photonic properties of one-dimensional systems have been extensively analyzed with this approach, not only because transfer matrix methods can be applied but also because such systems can be relatively easily produced in reality. There are examples based upon the Fibonacci sequence, the Thue-Morse sequence or Cantor sequences \cite{JPhysD.2007.Steurer,PhysRevLett.1987.Kohmoto,PhysRevE.2009.Esaki,JPhys.2007.Yin}, and also systems for negative refractive indices have been studied \cite{JPhys.1998.Vasconcelos,PhysLettA.2004.Li,JPhys.2006.deMedeiros}. Further, one can find a close resemblance of the theoretical and the experimental results \cite{PhysRevLett.1994.Gellermann,JPhys.2009.Nava}.

\begin{figure}
 \centering
 \includegraphics[width=8.8cm]{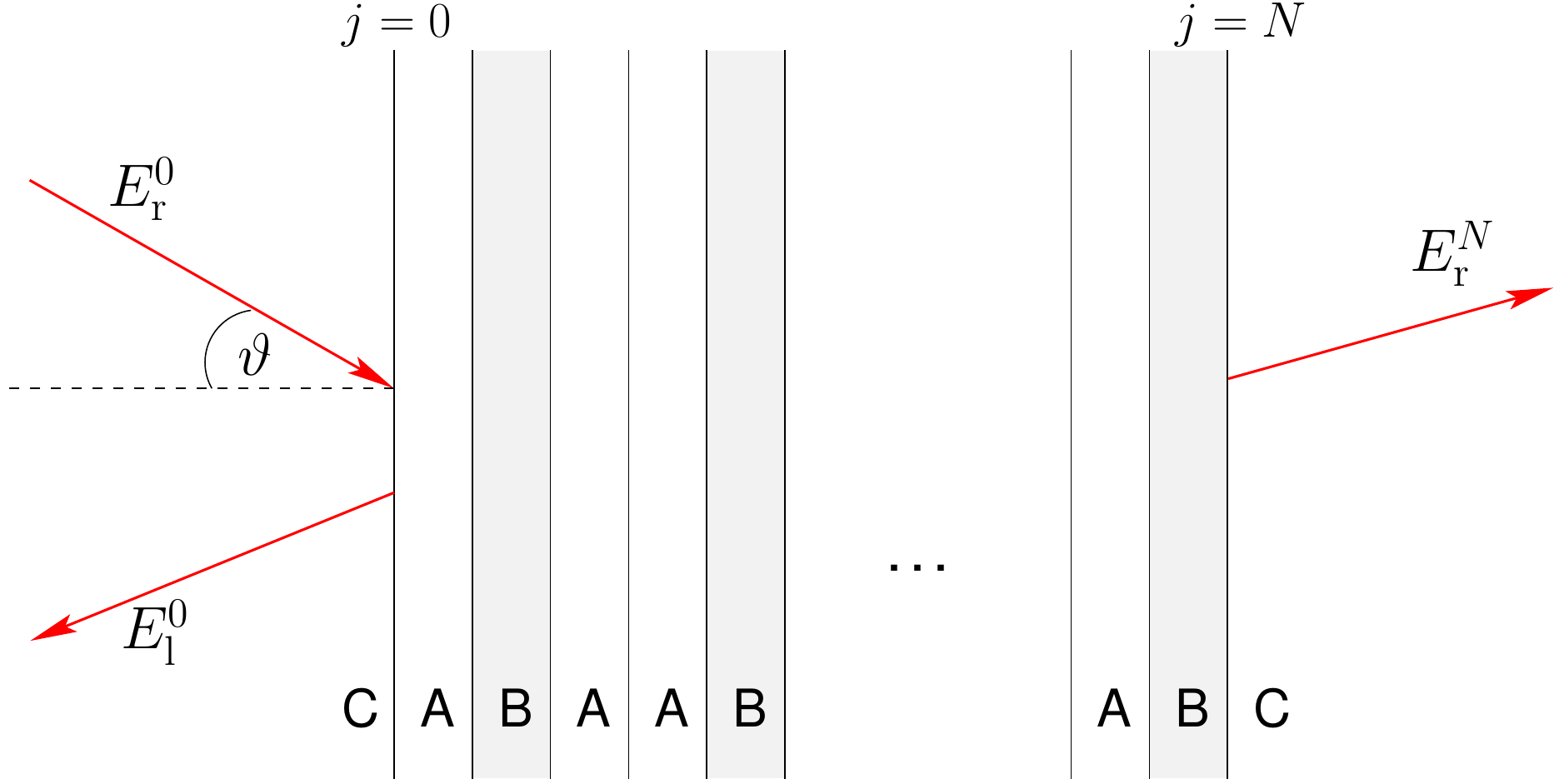}
 \caption{Transmission of light through a quasiperiodic stack consisting of the materials $A$ and $B$ with different indices of refraction. The light wave $E_\mathrm{r}^0$ is incident on the surface with an angle $\vartheta$. Parts of the light wave are transmitted through the stack ($E_\mathrm{r}^N$) and other parts are reflected ($E_\mathrm{l}^0$).}
 \label{fig:geometry}
\end{figure}

However, most of the investigations do not deal with the angle of incidence, but rather assume that the incident light wave is perpendicular to the layers of the stack, and do not investigate the influence of the structure of the quasiperiodic sequence on the transmission. In this paper we discuss the photonic properties of quasiperiodic systems consisting of a sequence of layers made of two different refractive indices $n_A$ and $n_B$ (cp.~Figure \ref{fig:geometry}), where the configuration of the layers is given according to the so-called metallic-mean sequences. In particular, we focus on the transmission of the light wave in dependence on the incidence angle $\vartheta$ and investigate the influence of the underlying construction rule for different metallic-mean quasicrystals on the photonic properties as well. We find that additional transmission bands occur for frequencies corresponding to the photonic band gap of the periodic system. Further, for small incident angles we obtain almost complete transmission for the quasiperiodic systems for a wide range of ratios of refractive indices $n_A/n_B$ in contrast to a system with periodically stacked layers, and for angles close to the angle of total reflection there is also a region of relatively high transmission.

The configuration of the layers of these so-called metal\-lic-mean quasicrystals can be constructed by the inflation rule
 \begin{equation}
  \label{equ:infaltion_rule}
  \mathcal{P}_a =
  \begin{cases}
    B \rightarrow A \\
    A \rightarrow ABA^{a-1}
  \end{cases}\hspace{1cm},
 \end{equation}
iterated $m$ times starting from the symbol $B$. For instance, this yields for the parameters $a=1,2,3$ the approximants
 \begin{align}
  B &\stackrel{\mathcal{P}_1}{\longrightarrow} A \stackrel{\mathcal{P}_1}{\longrightarrow} AB \stackrel{\mathcal{P}_1}{\longrightarrow}  ABA \stackrel{\mathcal{P}_1}{\longrightarrow} ABAAB \stackrel{\mathcal{P}_1}{\longrightarrow} \ldots \nonumber\\
  B &\stackrel{\mathcal{P}_2}{\longrightarrow} A \stackrel{\mathcal{P}_2}{\longrightarrow} ABA \stackrel{\mathcal{P}_2}{\longrightarrow}  ABAAABA \stackrel{\mathcal{P}_2}{\longrightarrow}  \ldots \\
  B &\stackrel{\mathcal{P}_3}{\longrightarrow} A \stackrel{\mathcal{P}_3}{\longrightarrow} ABAA \stackrel{\mathcal{P}_3}{\longrightarrow}  ABAAAABAAABAA \stackrel{\mathcal{P}_3}{\longrightarrow}  \ldots \nonumber\;.
 \end{align}
We refer to the resulting sequence after $m$ iterations as the $m$th order approximant $\mathcal{C}_m^a$ with the length $f_m$ given by the recursive equation $f_m = f_{m-2} + a f_{m-1}$ and $f_0 = f_1 = 1$. Depending on the parameter $a$ the inflation rule generates different metallic means, i.e., the lengths of two successive sequences satisfy the relation
 \begin{equation}
  \lim_{m \rightarrow \infty} \frac{f_m}{f_{m-1}} = \tau(a) \;,
 \end{equation}
where $\tau(a)$ is an irrational number with the continued fraction representation $\tau(a) = [\bar{a}] = [a,a,a,...]$. Also, the number $f_m^A$ of layers $A$ and number $f_m^B$ of layers $B$ in an approximant $\mathcal{C}_m^a$ are related by these metallic means according to
 \begin{equation}
  \label{equ:tau}
  \lim_{m \rightarrow \infty} \frac{f_m^A}{f_m^B} = \tau(a) \;.
 \end{equation}
For example, $a=1$ yields the well-known Fibonacci sequence related to the golden mean $\tau_\mathrm{Au} = [\bar{1}] = (1+\sqrt{5})/2$,
the case $a=2$ corresponds to the octonacci sequence with the silver mean $\tau_\mathrm{Ag} = [\bar{2}] = 1+\sqrt{2}$, and for $a=3$ one obtains the bronze mean $\tau_\mathrm{Bz} = [\bar{3}] = (3+\sqrt{13})/2$ \cite{JPhys.2007.Yin,PhysRevB.2009.Thiem}. In general the relation $\tau(a) = (a+\sqrt{a^2+4})/2$ holds \cite{NonLinAnal.1999.Spinadal}.

Due to the recursive inflation rule \eqref{equ:infaltion_rule}, these quasiperiodic chains possess a hierarchical structure, which is more
clearly visible by using the alternative construction rule
 \begin{equation}
 \label{equ:inflation_rule2}
 \mathcal{C}_m^a = \mathcal{C}_{m-1}^a\, \mathcal{C}_{m-2}^a\, (\mathcal{C}_{m-1}^a)^{a-1}\;,
 \end{equation}
yielding the same quasiperiodic sequences for $m \ge 2$ setting $\mathcal{C}_0^a = B$ and $\mathcal{C}_1^a = A$. Likewise, there is a corresponding expression for the recursive construction of the transfer matrix as given in Sec.~\ref{sec:transfer_matrix}.

The outline of this paper is as follows: In Sec.~\ref{sec:transfer_matrix} we give an introduction to the transfer matrix method used for the calculations and in Sec.~\ref{sec:results} we present our results and discuss them. This is followed by a brief conclusion in Sec.~\ref{sec:conclusion}.

\section{Transfer Matrix Method}\label{sec:transfer_matrix}

The propagation of light through a layered system as shown in Figure \ref{fig:geometry} is commonly investigated by means of transfer matrix methods \cite{PhysRevLett.1987.Kohmoto,PhysRevE.2009.Esaki,JPhys.1998.Vasconcelos,JPhys.2006.deMedeiros}. Regarding the geometry we consider the propagation of linearly polarized light with the electric field perpendicular to the plane of the light path (transverse electric waves) \cite{Book.Hecht}.

Within one layer the electric field $\mathbf{E} = E \mathbf{\hat{e}}_y$ can be described as a superposition of a right and a left traveling plane wave
\begin{equation}
 E = E_\mathrm{r}^j e^{ik_jx - i\omega t} + E_\mathrm{l}^j e^{-ik_jx - i\omega t} \;,
\end{equation}
where $k_j = n_j k$ denotes the wave number of the light in the $j$th layer made either of medium $A$ or $B$ with refractive index $n_j$, $k$ denotes the wave number in the vacuum and $\omega$ denotes the frequency of the light. Note, that each amplitude $E_\mathrm{r}^j$ and $E_\mathrm{l}^j$ actually represents the resultant of all possible waves traveling in the particular direction.

\begin{figure}
 \centering
 \includegraphics[width=0.75\columnwidth]{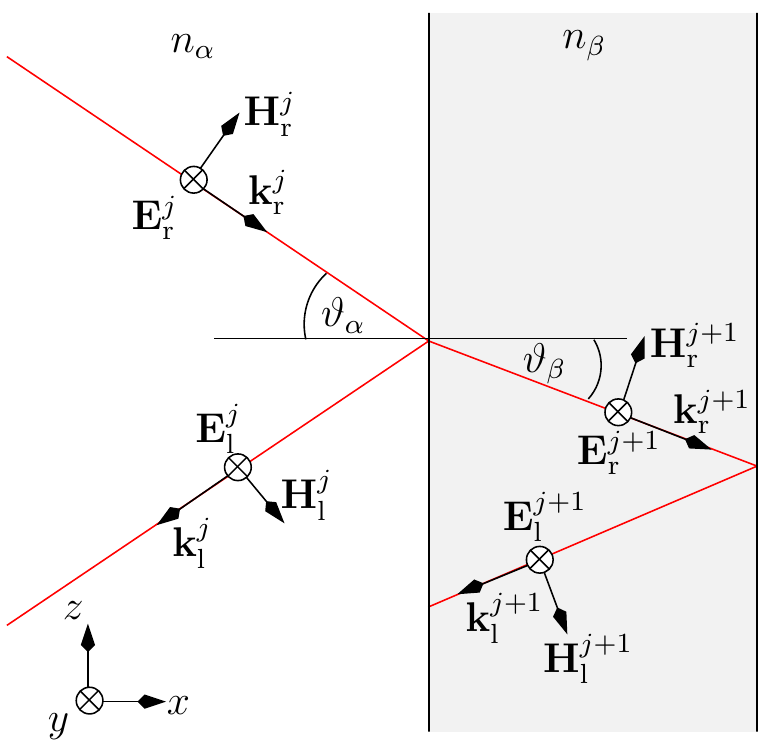}
 \caption{Electric and magnetic field vectors $\mathbf{E}$ and $\mathbf{H}$ at the interface between layer $j$ and $j+1$ with different indices of refraction $n_\alpha$ and $n_\beta$.}
 \label{fig:geometry2}
\end{figure}

The propagation of the waves through the stack is given on the one hand by their refraction at the interfaces between the layers and on the other hand by their propagation through the layers. Addressing the first point, the boundary conditions at the interfaces between the layers require the tangential component of the electric field $\mathbf{E}$ and the magnetic field $\mathbf{H} = \tfrac{n}{c} \mathbf{\hat{k}} \times \mathbf{E}$ to be continuous across the boundaries. This yields for an incidence angle $\vartheta_{\alpha}$, the emergence angle $\vartheta_{\beta}$, and the corresponding indices of refraction $n_\alpha$ and $n_\beta$ the relations (cp.~Figure \ref{fig:geometry2})
\begin{align}
 E_\mathrm{r}^j + E_\mathrm{l}^j &= E_\mathrm{r}^{j+1} + E_\mathrm{l}^{j+1} \\
 (E_\mathrm{r}^j - E_\mathrm{l}^j) n_\alpha \cos{\vartheta_\alpha} &= (E_\mathrm{r}^{j+1} - E_\mathrm{l}^{j+1}) n_\beta \cos{\vartheta_\beta} \;.
\end{align}
Applying the notation used by Kohmoto et al. \cite{PhysRevLett.1987.Kohmoto,PhysRevE.2009.Esaki} with the variables
\begin{equation}
 E_+ = E_\mathrm{r} + E_\mathrm{l} \text{  and  } E_- = -i\left(E_\mathrm{r} - E_\mathrm{l}\right) \;,
\end{equation}
the transfer of the light wave from medium $\alpha$ to medium $\beta$ can be described by
\begin{equation}
 \label{equ:transfer_equation}
 \begin{pmatrix}
  E_+ \\ E_-
 \end{pmatrix}_j = T_{\alpha\beta}
 \begin{pmatrix}
  E_+ \\ E_-
 \end{pmatrix}_{j+1}
\end{equation}
with the interface matrix
\begin{equation}
 T_{\alpha\beta} = T_{\beta\alpha}^{-1} =
 \begin{pmatrix}
  1 & 0 \\
  0 & \tfrac{n_{\beta} \cos{\vartheta_{\beta}}}{n_{\alpha} \cos{\vartheta_{\alpha}}}
 \end{pmatrix}\hspace{1cm}.
\end{equation}
Thereby, the incidence and emergence angles are related by Snell's law \cite{PhysRevLett.1987.Kohmoto}
\begin{equation}
 \label{equ:snell}
 \frac{\sin{\vartheta_{\alpha}}}{\sin{\vartheta_{\beta}}} = \frac{n_{\beta}}{n_{\alpha}} \;.
\end{equation}

\begin{figure}
 \centering
 \includegraphics[width=0.9\columnwidth]{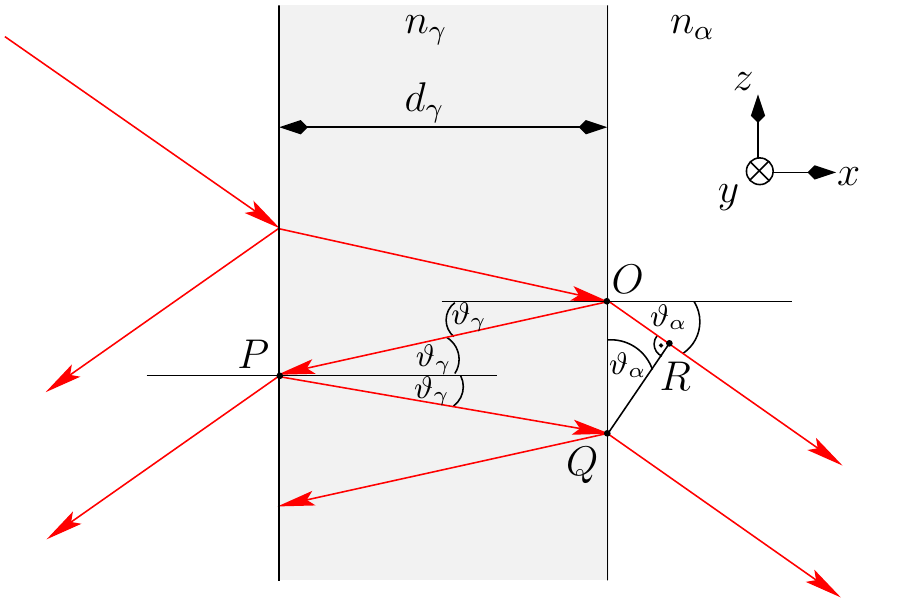}
 \caption{Phase difference between two waves originating from traversing a layer with index of refraction $n_\gamma$ and thickness $d_\gamma$.}
 \label{fig:geometry3}
\end{figure}

On the other hand, the propagation of the light waves within the layers results in a phase difference, which is comprised by the propagation matrix
\begin{equation}
 T_{\gamma} =
 \begin{pmatrix}
  \cos{\varphi_{\gamma}}& \sin{\varphi_{\gamma}} \\
  -\sin{\varphi_{\gamma}} & \cos{\varphi_{\gamma}}
 \end{pmatrix}\;.
\end{equation}
The phase difference for the wave length $\lambda = 2\pi / k$ of the light and a layer thickness $d_\gamma$ amounts to (cp. Figure \ref{fig:geometry3}) \cite{Book.Hecht}
\begin{equation}
 \label{equ:phase1}
 \varphi_{\gamma} = k [ n_\gamma ( \overline{OP} + \overline{PQ} ) - n_\alpha \overline{OR} ] = \frac{4\pi}{\lambda} d_{\gamma} n_{\gamma}\cos{\vartheta_{\gamma}} \;.
\end{equation}
Note, that the paper by Kohmoto et al. \cite{PhysRevLett.1987.Kohmoto} and some other related papers as e.g. \cite{JPhys.2007.Yin,JPhys.2009.Nava} only state the phase shift, which the waves undergo by traversing the stack, and not the phase difference between adjacent waves.

The transfer matrix
\begin{equation}
 \label{equ:transfer_equation2}
 \begin{pmatrix}
  E_+ \\ E_-
 \end{pmatrix}_0 = M_m^{\mathcal{C}^a}
 \begin{pmatrix}
  E_+ \\ E_-
 \end{pmatrix}_N
\end{equation}
of the system is then given as a combination of the different interface and propagation matrices according to a certain quasiperiodic sequence. Thereby, the amplitude of the incident light is denoted by $E_\mathrm{r}^{0}$, of the reflected light by $E_\mathrm{l}^{0}$ and of the transmitted light by $E_\mathrm{r}^{N}$ (cp.~Figure \ref{fig:geometry}).

For instance for the Fibonacci sequence $ABAAB \ldots B$ one obtains
\begin{multline}
 \hspace{0.5cm} M^{\mathcal{C}^\mathrm{Au}} = T_{CA}  T_A T_{AB} T_B T_{BA}T_A T_A T_{AB}T_B \\ \ldots \, T_B T_{BA} T_{AC} \;. \hspace{1.5cm}
\end{multline}
In general the recursive equation
\begin{equation}
 M_m^{\mathcal{C}^a} = M_{m-1}^{\mathcal{C}^a} M_{m-2}^{\mathcal{C}^a} \{ M_{m-1}^{\mathcal{C}^a} \}^{a-1}
\end{equation}
is applicable for $m \ge 2$ with $M_0^{\mathcal{C}^a} = T_{AB} T_B T_{BA} $ and $M_1^{\mathcal{C}^a} = T_A $, which has the same structure as the inflation rule of equation \eqref{equ:inflation_rule2}. Here we assume that either the surrounding medium has the same index of refraction as medium $A$ or one has to add the corresponding transfer matrices $T_{CA}$ and $T_{AC}$ at the edges, respectively.

In particular, we are interested in the calculation of the transmission coefficient $T$ (also known as transmittance) of the light through the stack, which is defined as $T = |E_\mathrm{r}^N|^2 / |E_\mathrm{r}^0|^2$ and can be derived from equation \eqref{equ:transfer_equation2} by eliminating $E_\mathrm{l}^0$. Assuming that there is no incident light from the right, i.e., $E_\mathrm{l}^{N}=0$, one obtains
\begin{equation}
 T = \frac{4}{|M|^2 + 2 \det{M}} \;,
\end{equation}
where $|M|^2$ corresponds to the sum of the squares of all four matrix elements of $M$. Further, for the complete stack it can be shown that $\det{M} = 1$, which simplifies the expression for the transmission coefficient $T$.

\section{Results}\label{sec:results}

\begin{figure*}
 \centering
 \includegraphics[width=8.7cm]{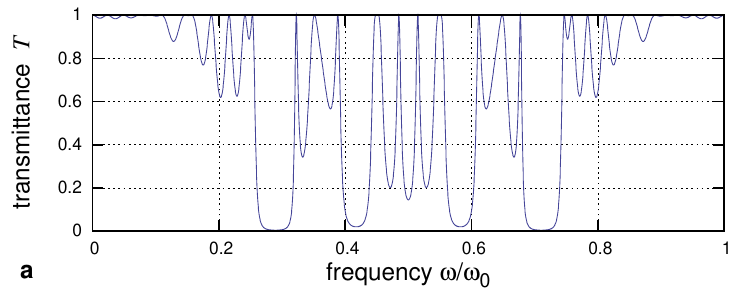}\hspace{0.2cm}
 \includegraphics[width=8.7cm]{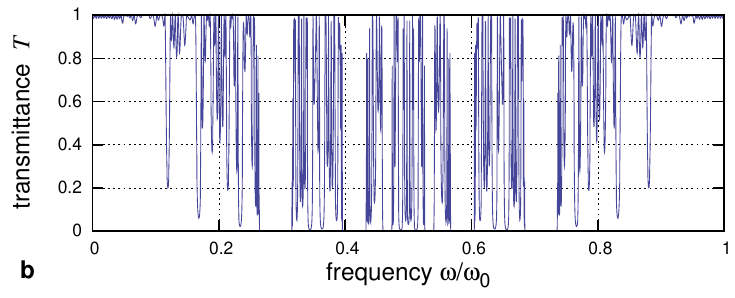}
 \includegraphics[width=8.7cm]{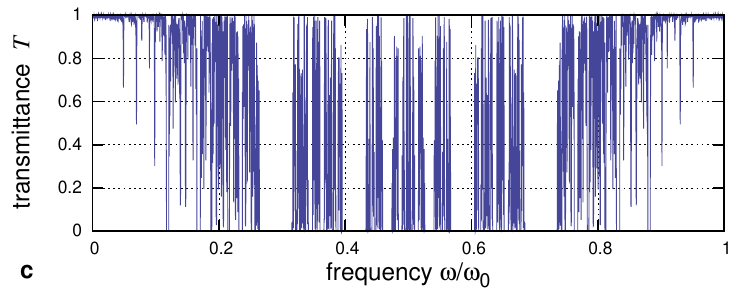}\hspace{0.2cm}
 \includegraphics[width=8.7cm]{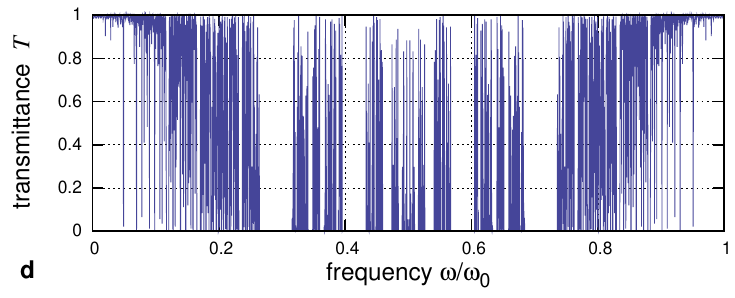}
 \caption{Transmittance $T(\omega)$ for different approximants of the octonacci chain with incidence angle $\vartheta = 0$ and indices of refraction $n_A = n_C$ and $u = n_A /n_B = 2/3$: (a) $\mathcal{C}_5^\mathrm{Ag}$ with $f_5 = 41$ layers, (b) $\mathcal{C}_7^\mathrm{Ag}$ with $f_7 = 239$ layers, (c) $\mathcal{C}_9^\mathrm{Ag}$ with $f_9 = 1393$ layers, and (d) $\mathcal{C}_{11}^\mathrm{Ag}$ with $f_{11} = 8119$ layers.}
 \label{fig:Tangle0}
\end{figure*}

In this section we comprise the results for several metallic-mean quasicrystals. Especially, we focus on the change of the transmission coefficient $T$ in dependence on the used inflation rule $\mathcal{P}$ and the incidence angle $\vartheta$.

For the choice of the wave length of the incident light $\lambda_0$, we use the commonly applied quarter wave length condition $n_A d_A = n_B d_B = \lambda_0/4$, which leads for an incidence angle $\vartheta = 0$ to an identical optical wave path for the light in the two materials $A$ and $B$ of the stack \cite{JPhys.2006.deMedeiros,PhysRevLett.1994.Gellermann,JPhys.2009.Nava}. Hence, equation \eqref{equ:phase1} results in
\begin{equation}
 \label{equ:phase2}
 \varphi_{\gamma} = \frac{\lambda_0}{\lambda} \pi \cos{\vartheta_{\gamma}} = \frac{\omega}{\omega_0} \pi \cos{\vartheta_{\gamma}} \;.
\end{equation}

\begin{figure*}
 \centering
 \includegraphics[width=0.9\columnwidth]{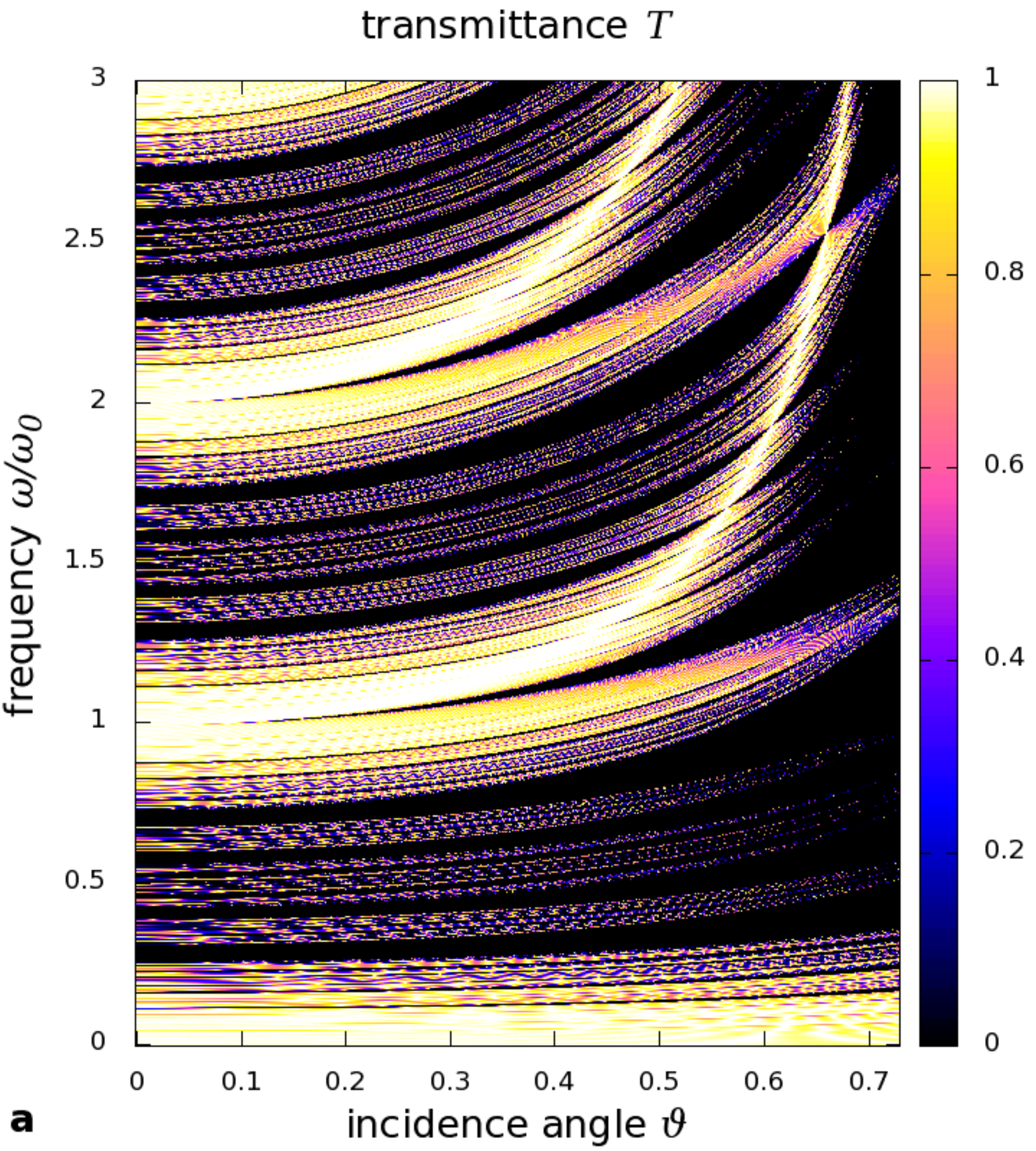}\hspace{0.5cm}
 \includegraphics[width=0.9\columnwidth]{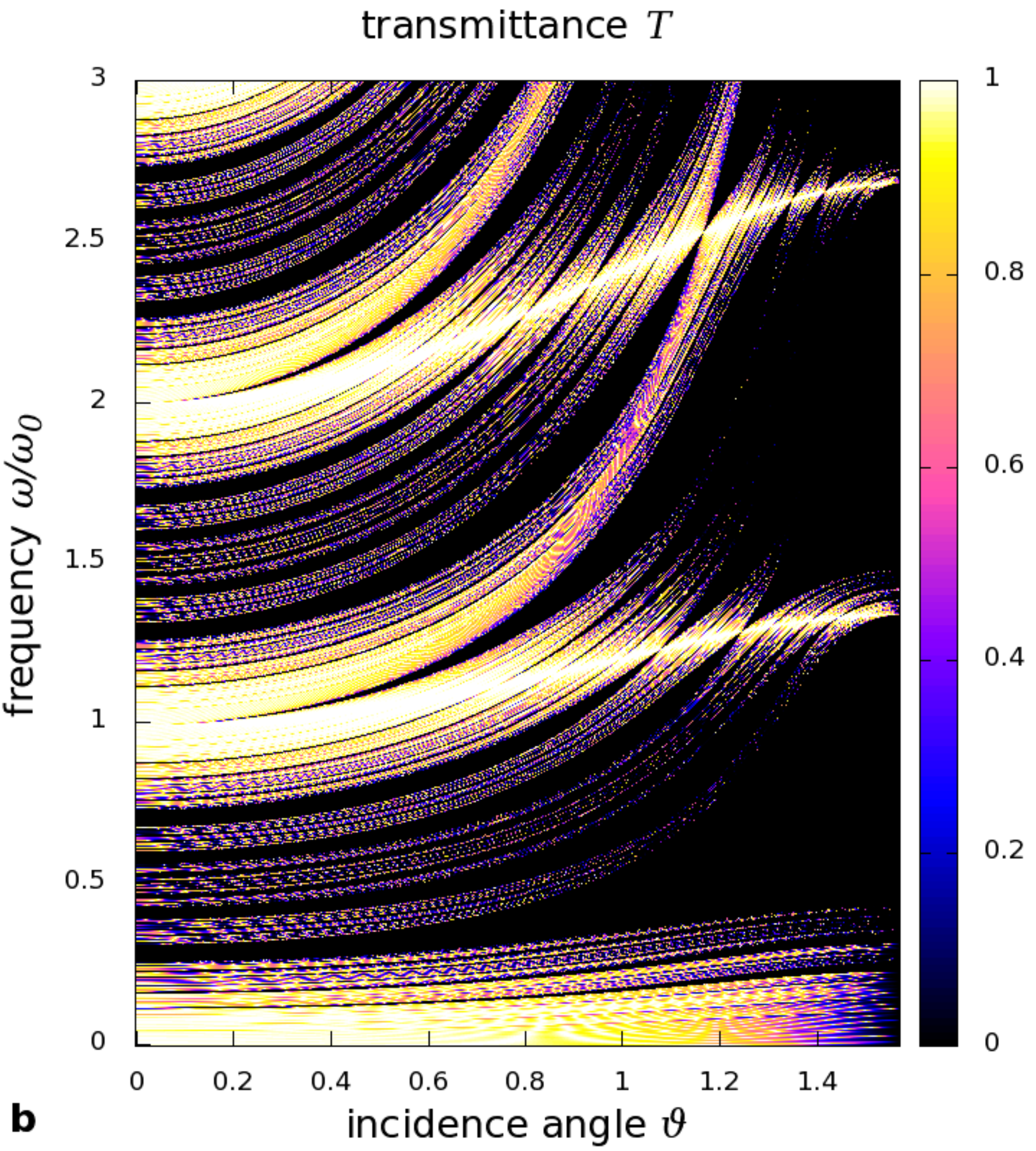}
 \caption{Transmittance $T(\vartheta, \omega)$ of light through a stack of the materials A and B, which are arranged according to the octonacci sequence $\mathcal{C}_{8}^{\mathrm{Ag}}$ with $f_8 = 577$ layers. Results are shown in dependence on the reduced frequency $\omega / \omega_0$ versus the incidence angle $\vartheta$ for $n_A = n_C$ and (a) $u = 3/2$ and (b) $u = 2/3$.}
 \label{fig:Toctonacci}
\end{figure*}

In Figure \ref{fig:Tangle0} the results for the transmission coefficients $T(\omega)$ are shown for different approximants of the octonacci chain $\mathcal{C}_m^{\mathrm{Ag}}$ for an incidence angle $\vartheta = 0$. Due to the structure of the interface and propagation matrices only the ratio of the indices of refraction $u = n_A/n_B$ is important. Here we choose a fixed ratio $u = 2/3$, and $n_C = n_A$ for the surrounding medium. Results are only displayed in the interval $[0,\omega/\omega_0]$ because the pattern of the transmittance $T$ repeats every $\omega/\omega_0$ for $\vartheta = 0$ (cp. Figure \ref{fig:Toctonacci}). In general, the transmission spectrum of metallic-mean sequences or Cantor sequences is assumed to show either complete transmission or complete reflectance for infinite systems \cite{PhysRevE.2009.Esaki,JPhysA.2006.Honda} similar to the singular continuous energy spectrum of quasiperiodic systems \cite{JPhysFrance.1989.Sire,JPhysA.1989.Gumbs,JStatPhys.1989.Suto}. This characteristic becomes more and more visible for larger system sizes $f_m$. However, especially for small approximants there still occurs a significant transmittance $T(\omega) > 0$ in the photonic band gaps since the waves can tunnel through the stack to a certain extent. Likewise, $T(\omega) = 1$ is often not reached within the bands.

When varying the incidence angle, we have to keep in mind that total reflection occurs for $n_B < n_A$ at the incidence angles $\vartheta > \vartheta_{\mathrm{total}} = \arcsin{ u^{-1}}$ and hence we can only expect to obtain transmission coefficients for incidence angles $\vartheta < \vartheta_{\mathrm{total}}$.

\begin{figure*}
 \centering
 \includegraphics[height=5.65cm]{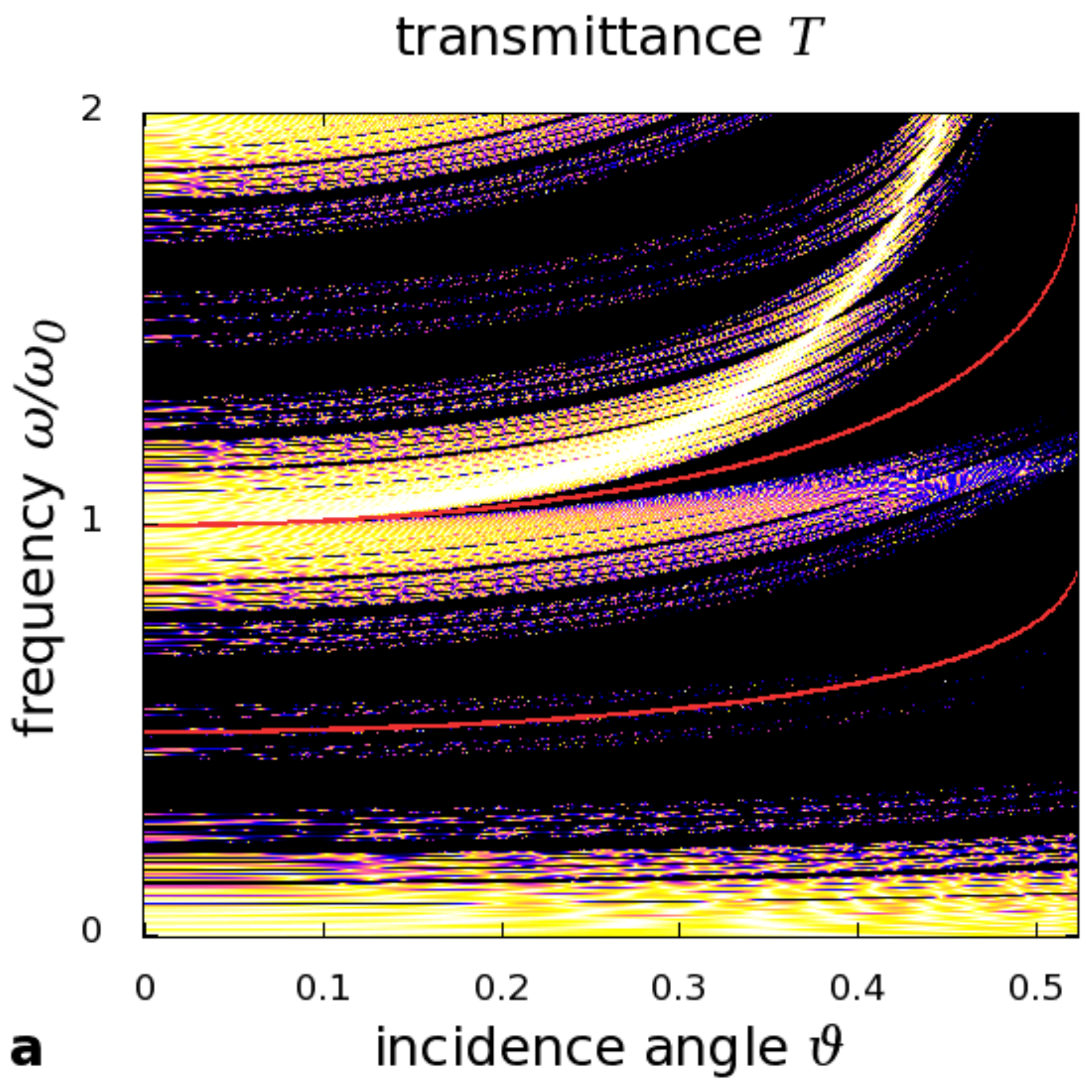}\hspace{0.1cm}
 \includegraphics[height=5.65cm]{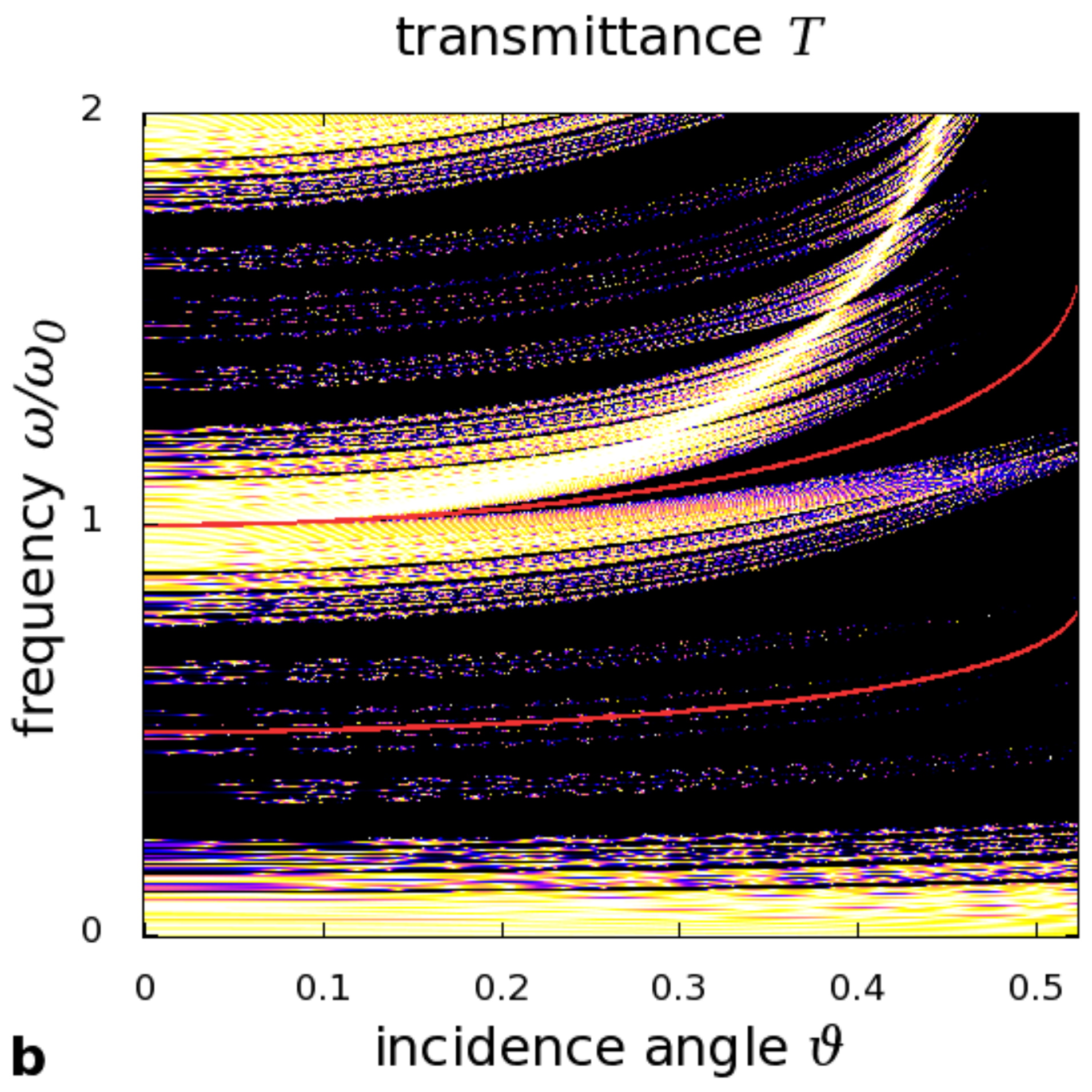}\hspace{0.1cm}
 \includegraphics[height=5.65cm]{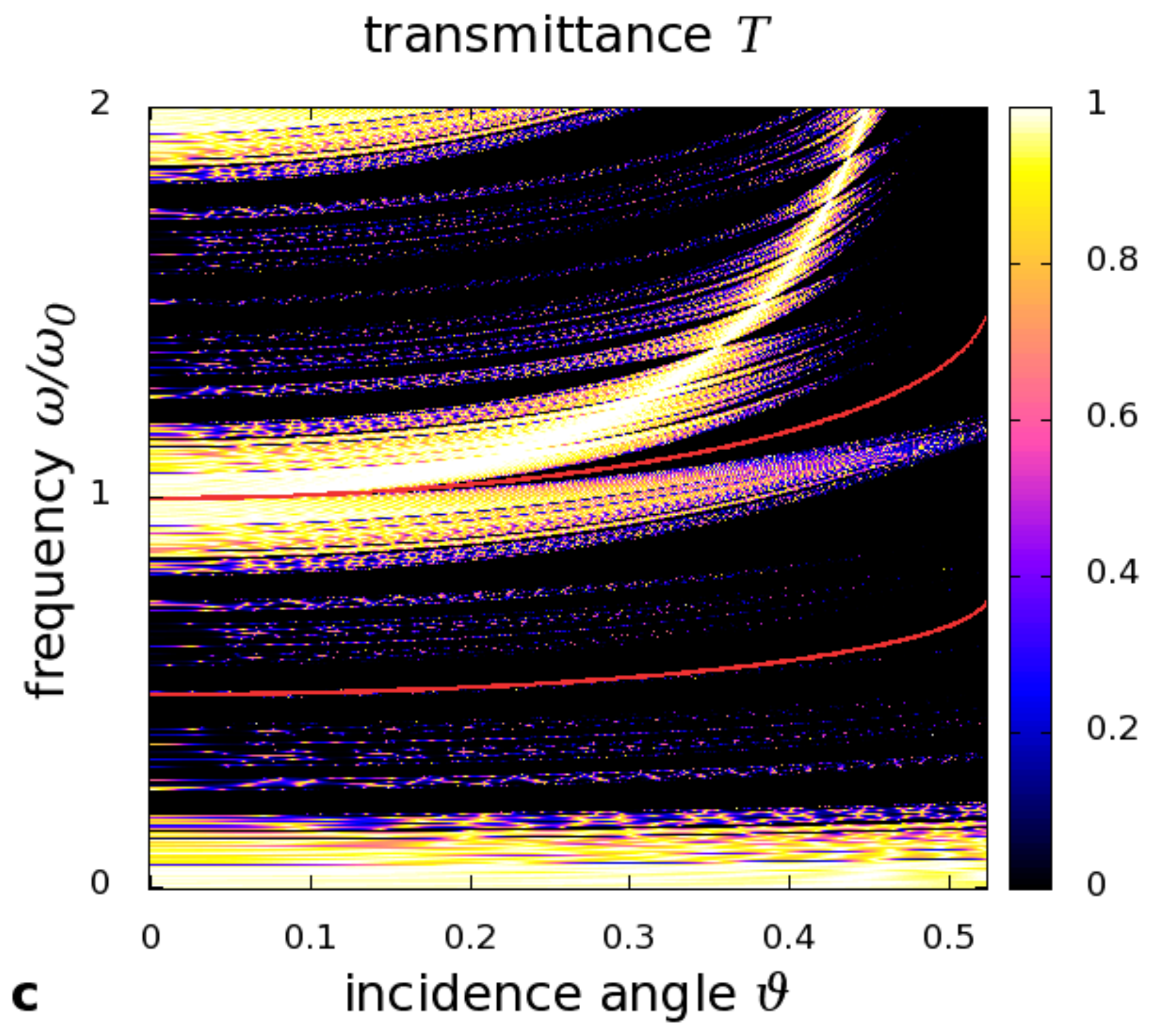}
 \caption{Transmittance $T(\vartheta, \omega)$ for different quasiperiodic inflation rules: (a) the Fibonacci sequence $\mathcal{C}_{14}^{\mathrm{Au}}$ with $f_ {14} = 610$ layers, (b) the octonacci sequence $\mathcal{C}_8^{\mathrm{Ag}}$ with $f_8 = 577$ layers and (c) the bronze mean quasiperiodic sequence $\mathcal{C}_6^{\mathrm{Br}}$ with $f_6 = 469$ layers. The red lines show the expected limit behavior of the bending according to equation \eqref{equ:bendingT4}. The ratio of the indices of refraction is $u=2$.}
 \label{fig:T_all_angles}
\end{figure*}

The transmittance $T(\vartheta, \omega)$ in dependence on the reduced frequency $\omega / \omega_0$ and the incidence angle $\vartheta < \vartheta_\mathrm{total}$ is shown in Figure \ref{fig:Toctonacci} and Figure \ref{fig:T_all_angles} for different metallic-mean quasicrystals. Figure \ref{fig:Toctonacci} displays the transmission for the 8th approximant of the octonacci sequence for a broad range of frequencies for two examples corresponding to the cases $n_A < n_B$ and $n_B > n_A$. As mentioned, for $\vartheta = 0$ the transmission spectrum is periodic with respect to the frequency. However, for all angles $\vartheta > 0$ the photonic transmission bands bend towards higher frequencies for increasing incidence angles and this effect becomes stronger for higher frequencies. Further, for $u > 1$ one obtains a relatively large range of frequencies with nearly complete transmission for large incidence angles and for $u < 1$ there are certain frequencies with nearly complete transmission over a large range of angles due to the bending of the transmission bands.

\begin{figure*}[t]
 \centering
 \includegraphics[height=7.3cm]{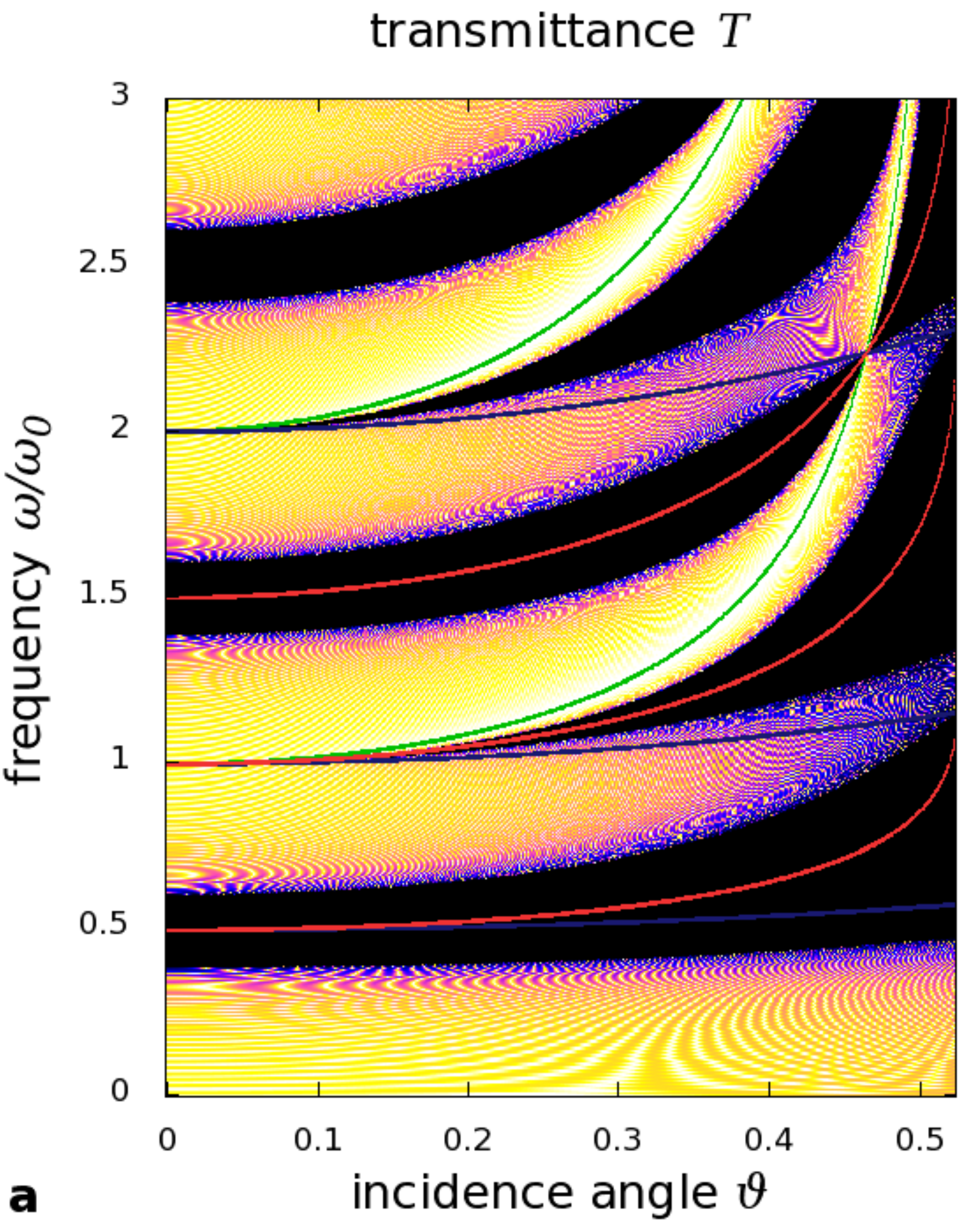}\hspace{0.2cm}
 \includegraphics[height=7.3cm]{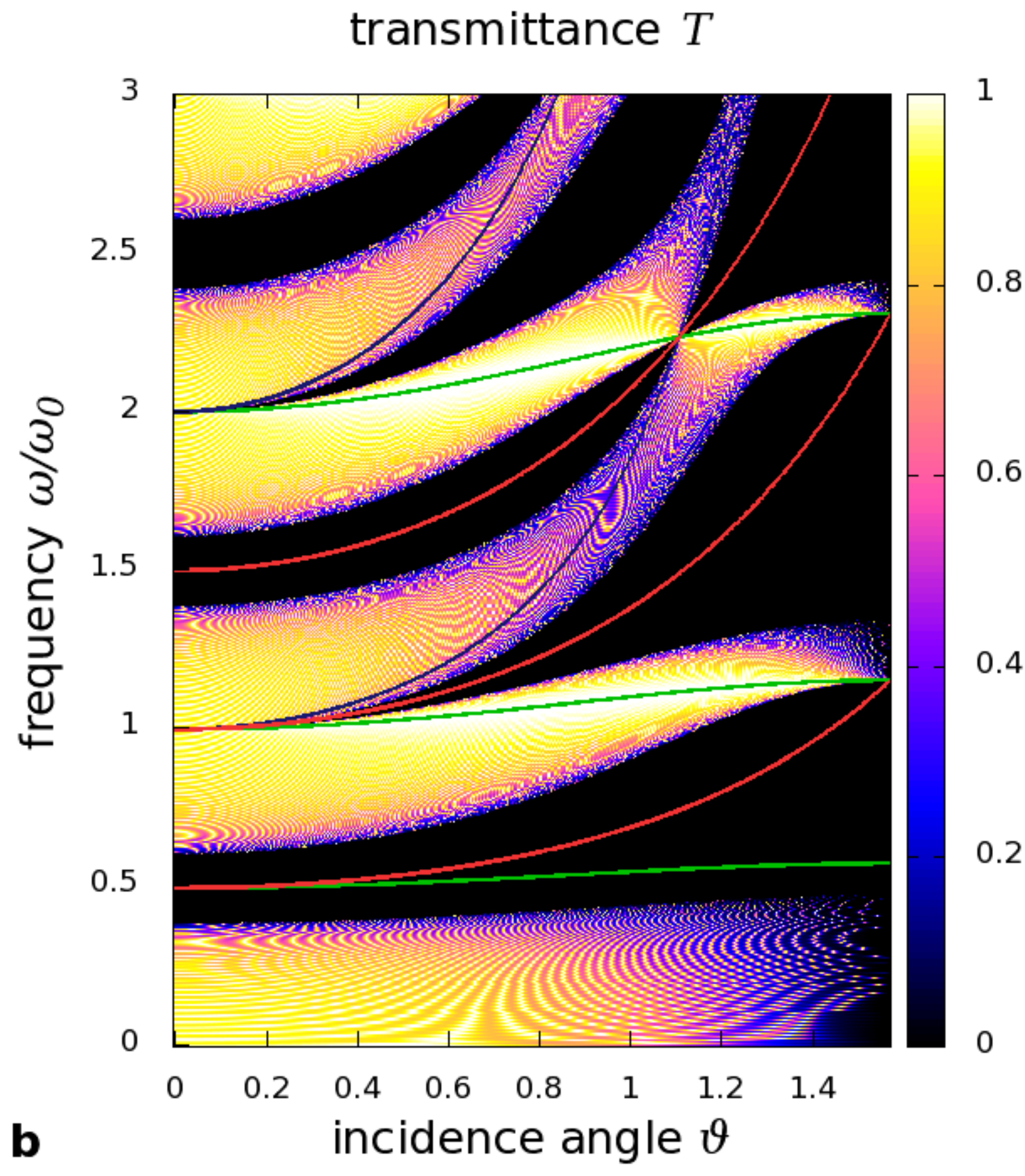}\hspace{0.2cm}
 \includegraphics[width=5.3cm]{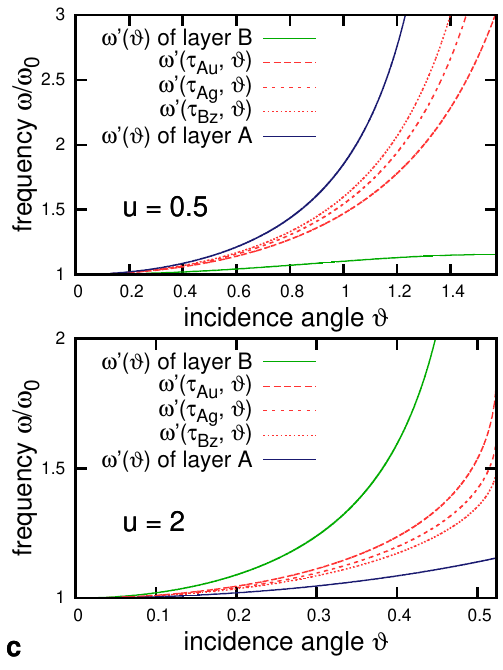}
 \caption{Transmittance $T(\vartheta, \omega)$ for a periodic sequence of alternating layers $ABAB \ldots$ with length $500$
 for the ratios of the indices of refraction (a) $u = 2$ as in Figure \ref{fig:T_all_angles} and (b) $u = 0.5$. The additional lines in (a) and (b) show the expected limit behavior of the bending: the green, blue, and red lines correspond to the equations \eqref{equ:bendingT1}, \eqref{equ:bendingT2}, and \eqref{equ:bendingT3} respectively. Plot (c) shows these limit functions as well but additionally the bending for quasiperiodic sequences in dependency on the metallic mean $\tau$ according to equation \eqref{equ:bendingT4}. The Moir\'e patterns in (a) and (b) are caused by strong oscillations of the transmittance $T$, which are not adequately displayed due to the used resolution.}
 \label{fig:bendingT}
\end{figure*}

In Figure \ref{fig:T_all_angles} the transmittance $T(\vartheta, \omega)$ is displayed for different metallic-mean sequences. Comparing with the results obtained for a periodic alignment of the layers $A$ and $B$ in Figure \ref{fig:bendingT}, one can see that there is a typical photonic band gap at $\omega = \omega_0/2 $ for the periodic chain for $\vartheta = 0$. Also for $\omega / \omega_0 = b+1/2$ ($b \in \mathbb{N}_0$) such gaps occur and with increasing incidence angles these photonic band gaps bend also towards higher frequencies. We denote the corresponding frequencies of these photonic band gaps as mid-gap frequencies. For the quasiperiodic sequences this band gap appears to be much wider, but is intersected by an increasing number of new lines of moderately high transmission for increasing $a$ of the inflation rule. For $\omega / \omega_0 = b$ ($b \in \mathbb{N}_0$) the characteristics of the transmittance $T(\vartheta, \omega)$ hardly change for the different construction rules. The behavior at the photonic band gap is in consistency with the observation that for $\omega = \omega_0/2$ ($\vartheta = 0$) the quasiperiodicity is most effective \cite{JPhysJap.1991.Kono2}. For $u > 1$ the narrow transmission bands caused by the quasiperiodicity do not bend as strongly as the transmission bands already present for the periodic case. Therefore, for large incidence angles $\vartheta$ the bands corresponding to the periodic case intersect the photonic band gaps and complete transmission occurs. In this case the transmission spectrum for the quasiperiodic and the periodic case look quite similar. For $u < 1$ all bands bend nearly equally strong but also form regions of almost complete transmission at certain frequencies for a large range of incidence angles (cp.~Figure \ref{fig:Toctonacci}(b)). Additionally, in both cases we obtain that the additional transmission bands are more uniformly distributed with increasing $a$ of the inflation rule and hence the photonic band gaps become smaller.

The dependency of the transmission bands on the incidence angle can be explained by the difference of the optical light paths in the layers $A$ and $B$ for $\vartheta > 0$. In this case we expect that the transmission coefficients $T(\omega, \vartheta = 0)$ and $T(\omega^\prime, \vartheta > 0)$ yield almost the same results if the overall phase difference for the stack is equal. At first, we consider only the behavior arising from the change of the phase difference within one layer. Hence, for layer $B$ the relation
\begin{equation}
 \varphi_B (\omega, \vartheta = 0) \stackrel{!}{=} \varphi_B (\omega^{\prime}, \vartheta_B > 0)
\end{equation}
has to be fulfilled, which yields a behavior according to
 \begin{equation}
  \label{equ:bendingT1}
  \omega^\prime (\vartheta) \stackrel{\eqref{equ:phase2}}{=} \frac{\omega (\vartheta = 0)}{ \cos{\vartheta_B} } \stackrel{\eqref{equ:snell}}{=} \frac{ \omega (\vartheta = 0)}{ \cos{\left(\arcsin{ \left(u \sin{\vartheta}\right)}\right)}} \;.
 \end{equation}
Note that for $n_A = n_C$ the relation $\vartheta_A = \vartheta$ holds. Likewise, the relation for layer $A$ can be derived with
 \begin{equation}
  \label{equ:bendingT2}
  \omega^\prime (\vartheta) \stackrel{\eqref{equ:phase2}}{=} \frac{\omega (\vartheta = 0)}{ \cos{\vartheta} }\;.
 \end{equation}
However, the actual bending of the transmission bands is given as a superposition of equations \eqref{equ:bendingT1} and \eqref{equ:bendingT2}. For a periodic stack with the same number of layers $A$ and $B$ the total phase difference $\varphi = (\varphi_A + \varphi_B)f_m/2$ results in a dependency
 \begin{equation}
  \label{equ:bendingT3}
  \omega^\prime (\vartheta) = \frac{2\omega (\vartheta = 0)}{ \cos{\vartheta} + \cos{\left(\arcsin{ \left(u \sin{\vartheta}\right)}\right)} }\;.
 \end{equation}
The functions \eqref{equ:bendingT1}--\eqref{equ:bendingT3} are shown in comparison with the numerical results of the transmission in a periodic stack of layers $A$ and $B$ in Figure \ref{fig:bendingT}(a) and (b). The functions \eqref{equ:bendingT3} fit the bending of the transmission bands quite well, especially for small angels. Further, we obtained that the regions of almost total transmission behave according to equation \eqref{equ:bendingT1}, which indicates that $\varphi_B$ plays a significant part for this behavior. In contradistinction the bending caused by $\varphi_A$ often coincides with regions of lower transmission coefficients. This effect becomes even more evident for systems with $u \ll 1$ or $u \gg 1$.

For the quasiperiodic stacks the numbers of layers $A$ and $B$ are not identical but related to the metallic means $\tau$ by equation \eqref{equ:tau}. The overall phase difference in the limit of infinite stacks is then given by
 \begin{equation}
 \varphi = \left(\frac{\tau}{1+\tau}\varphi_A + \frac{1}{1+\tau}\varphi_B \right)f_m
 \end{equation}
and yields a bending according to
 \begin{equation}
  \label{equ:bendingT4}
  \omega^\prime (\vartheta) = \frac{(1+\tau) \omega (\vartheta = 0) }{ \tau \cos{\vartheta} + \cos{\left(\arcsin{ \left(u \sin{\vartheta}\right)}\right)} }\;.
 \end{equation}
These functions are shown in Figure \ref{fig:bendingT}(c) for the two cases $u=0.5$ and $u=0.2$. Again, we obtain a bending in between the behavior arising from one single layer of medium $A$ or $B$. In particular, the bending increases with the parameter $\tau$ (respectively $a$) in the case $u < 1$ and shows the reverse behavior for $u > 1$. Further, for the quasiperiodic systems the bending according to equation \eqref{equ:bendingT4} approaches for $\tau \to 1$ the bending of a periodic stack and for $\tau \gg 1$ the behavior arising from one single layer $B$. Comparing the functions \eqref{equ:bendingT4} with the numerical results of the transmission for the quasiperiodic layered systems with different values of $\tau$ (cp. Figure \ref{fig:T_all_angles}), we find a very good resemblance of the bending behavior for small and intermediate angles.

\begin{figure*}
 \centering
 \includegraphics[height=5.65cm]{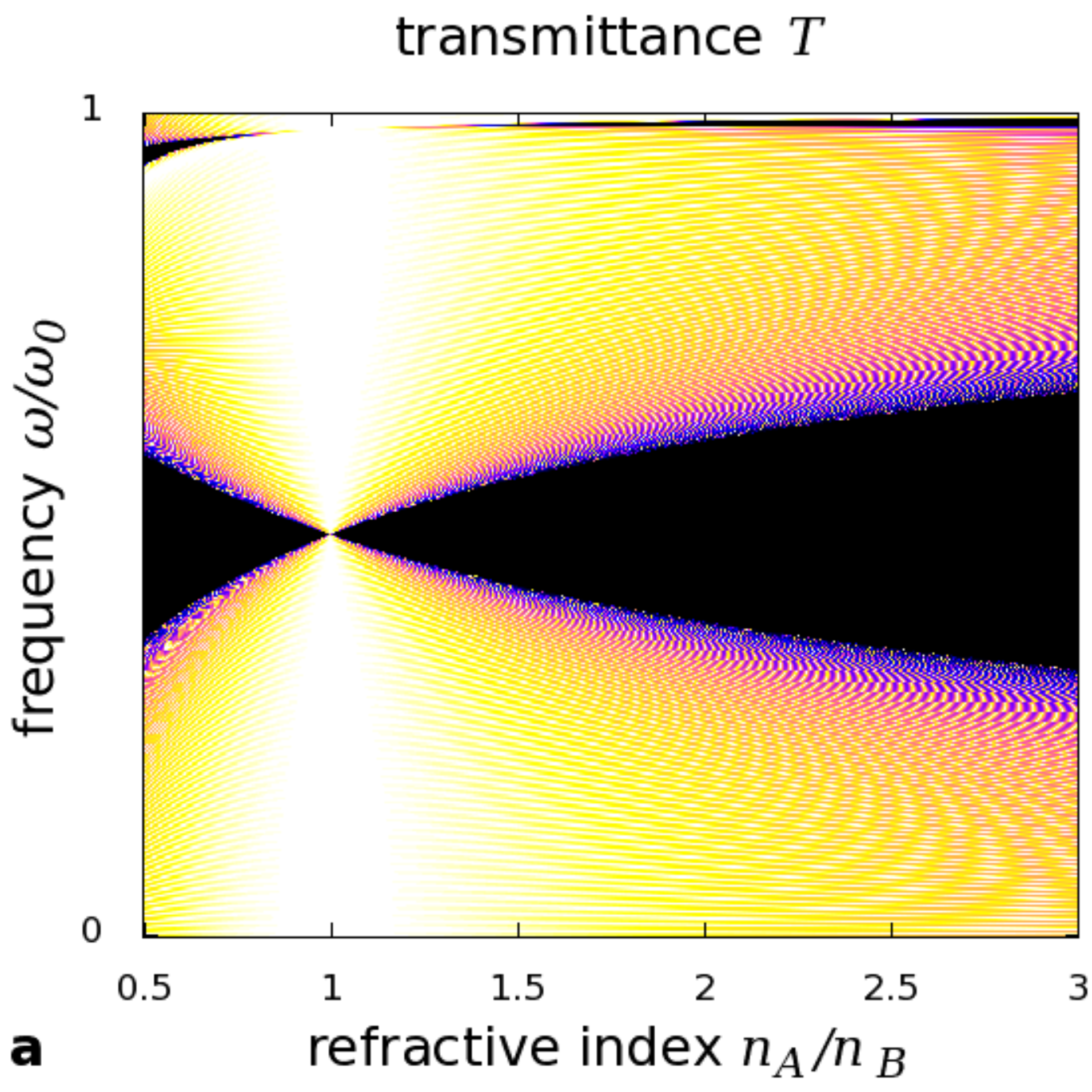}\hspace{0.1cm}
 \includegraphics[height=5.65cm]{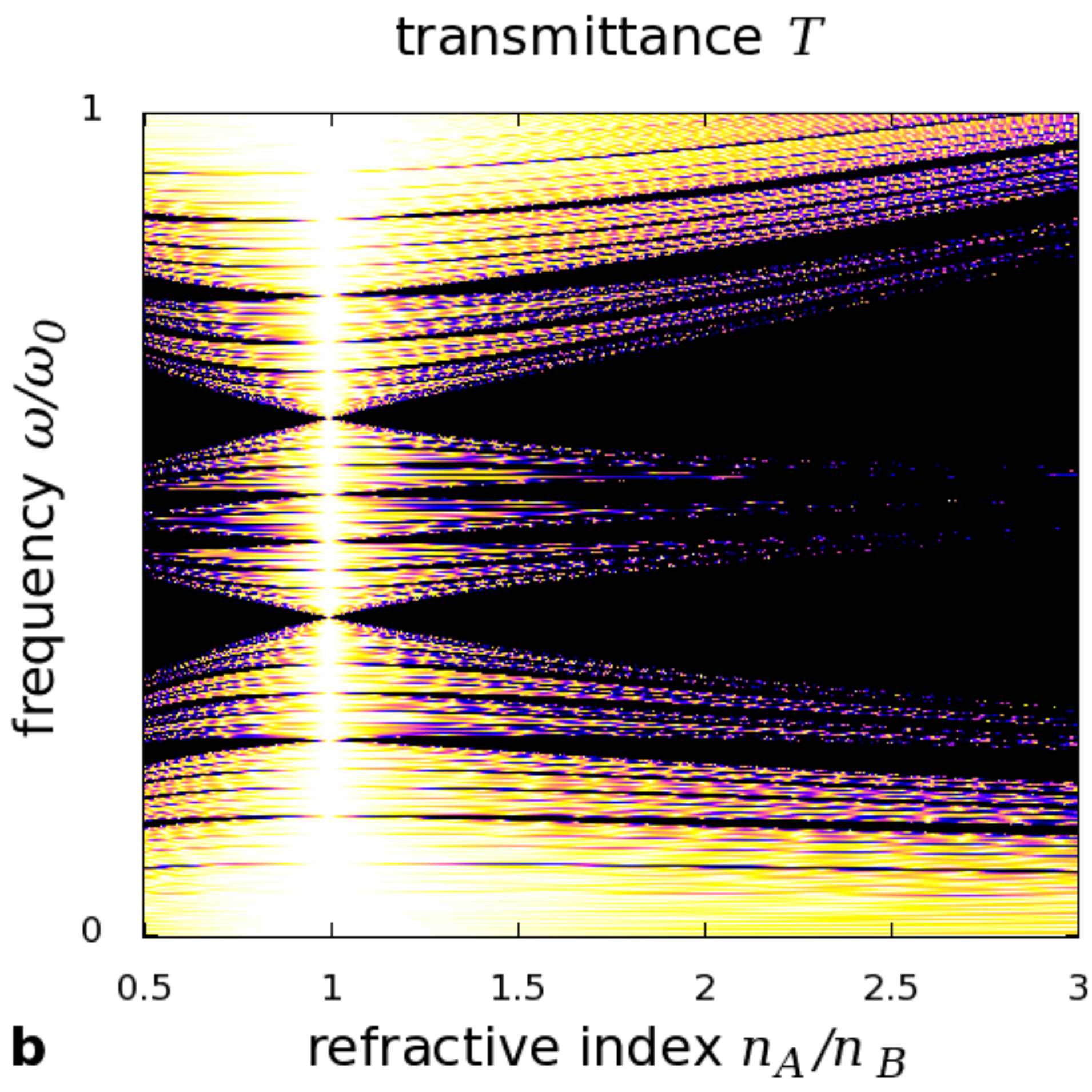}\hspace{0.1cm}
 \includegraphics[height=5.65cm]{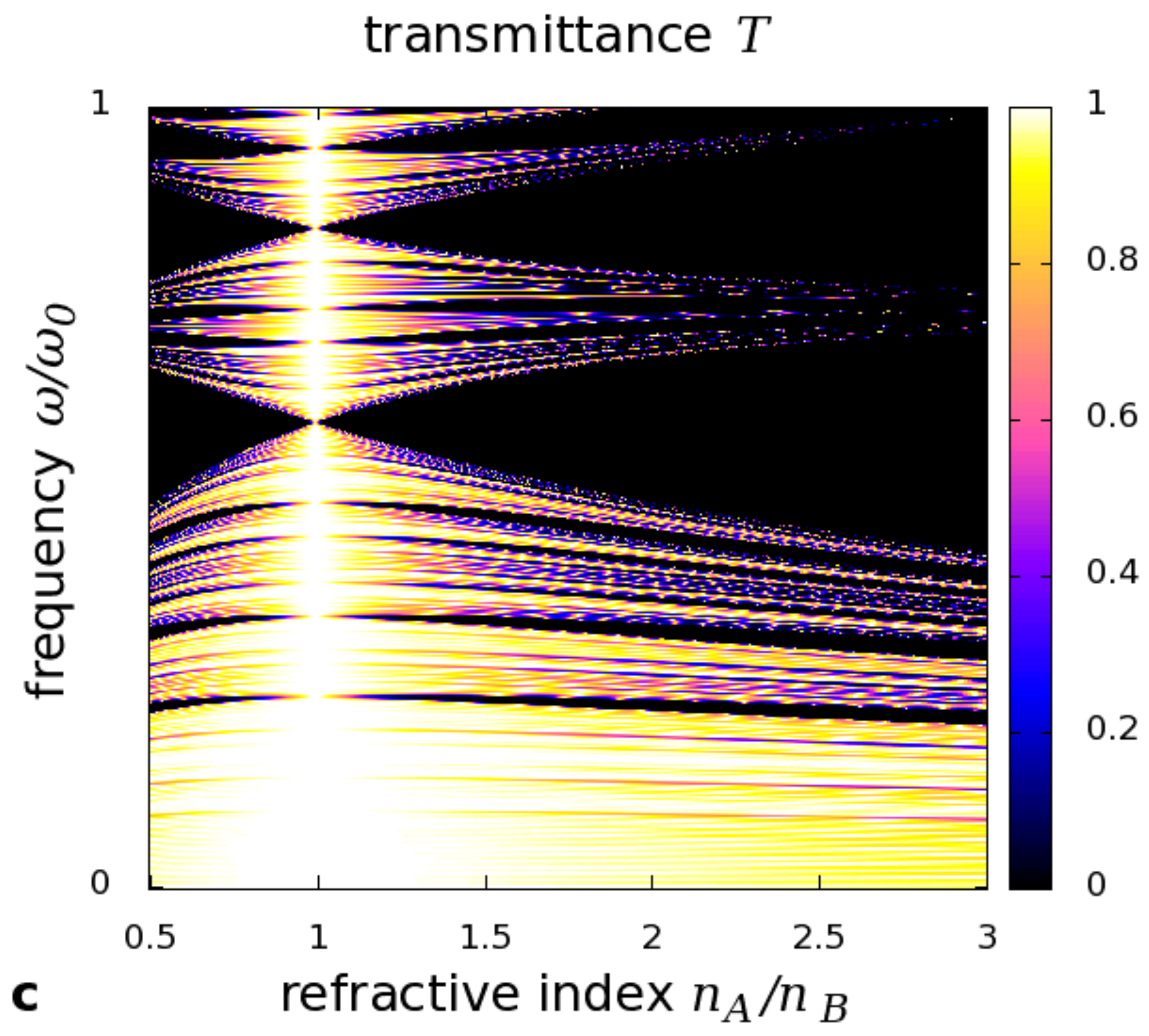}
 \includegraphics[height=5.65cm]{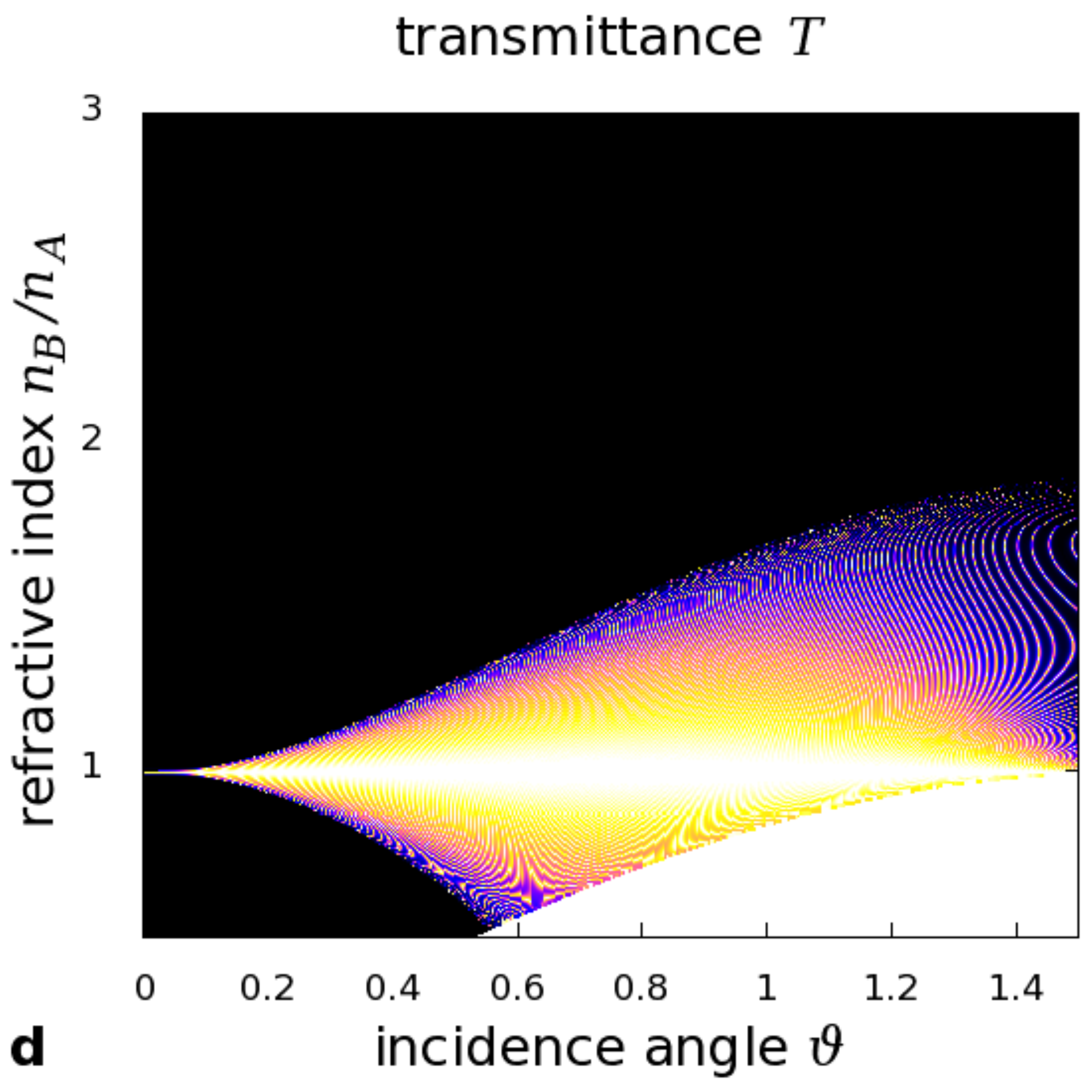}\hspace{0.1cm}
 \includegraphics[height=5.65cm]{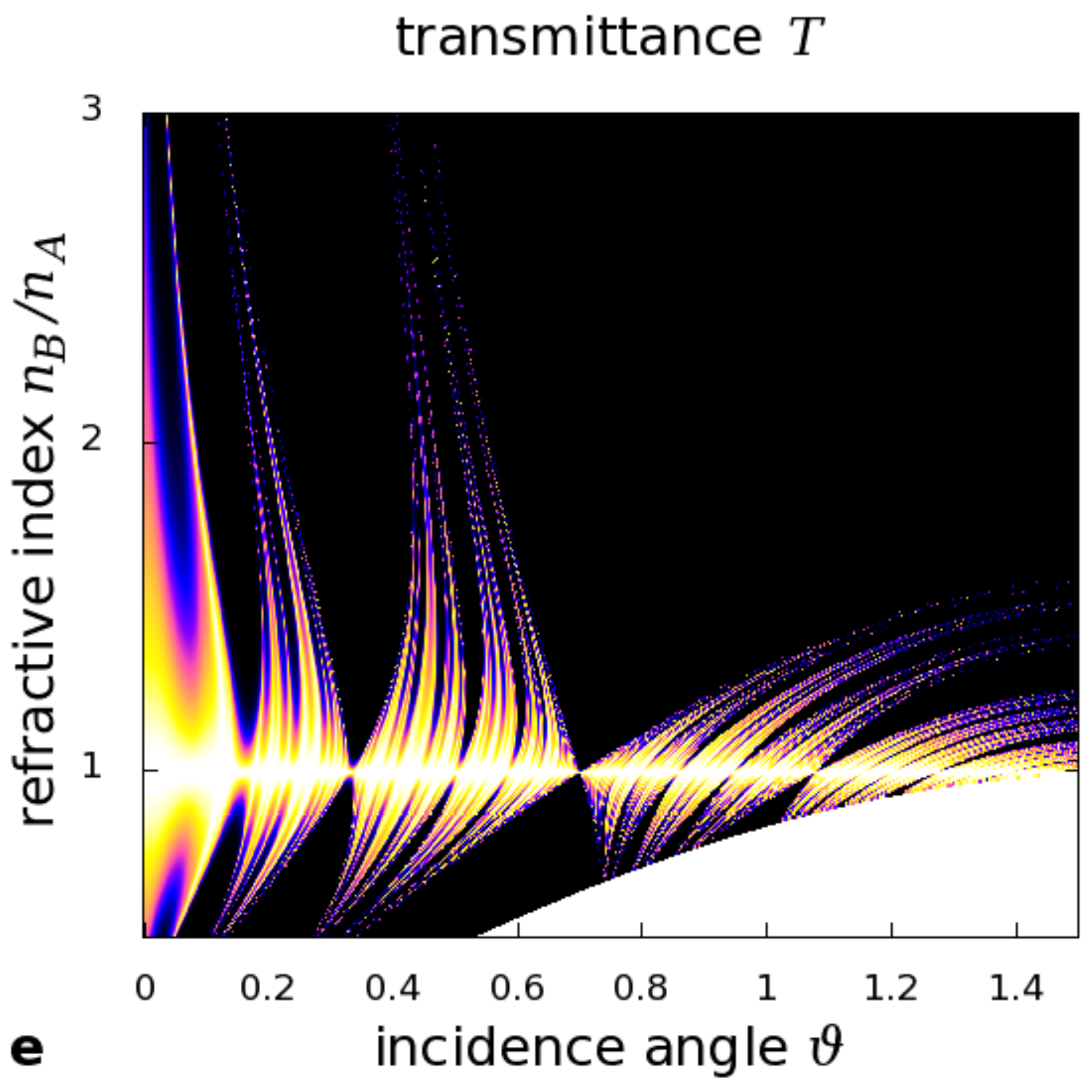}\hspace{0.1cm}
 \includegraphics[height=5.65cm]{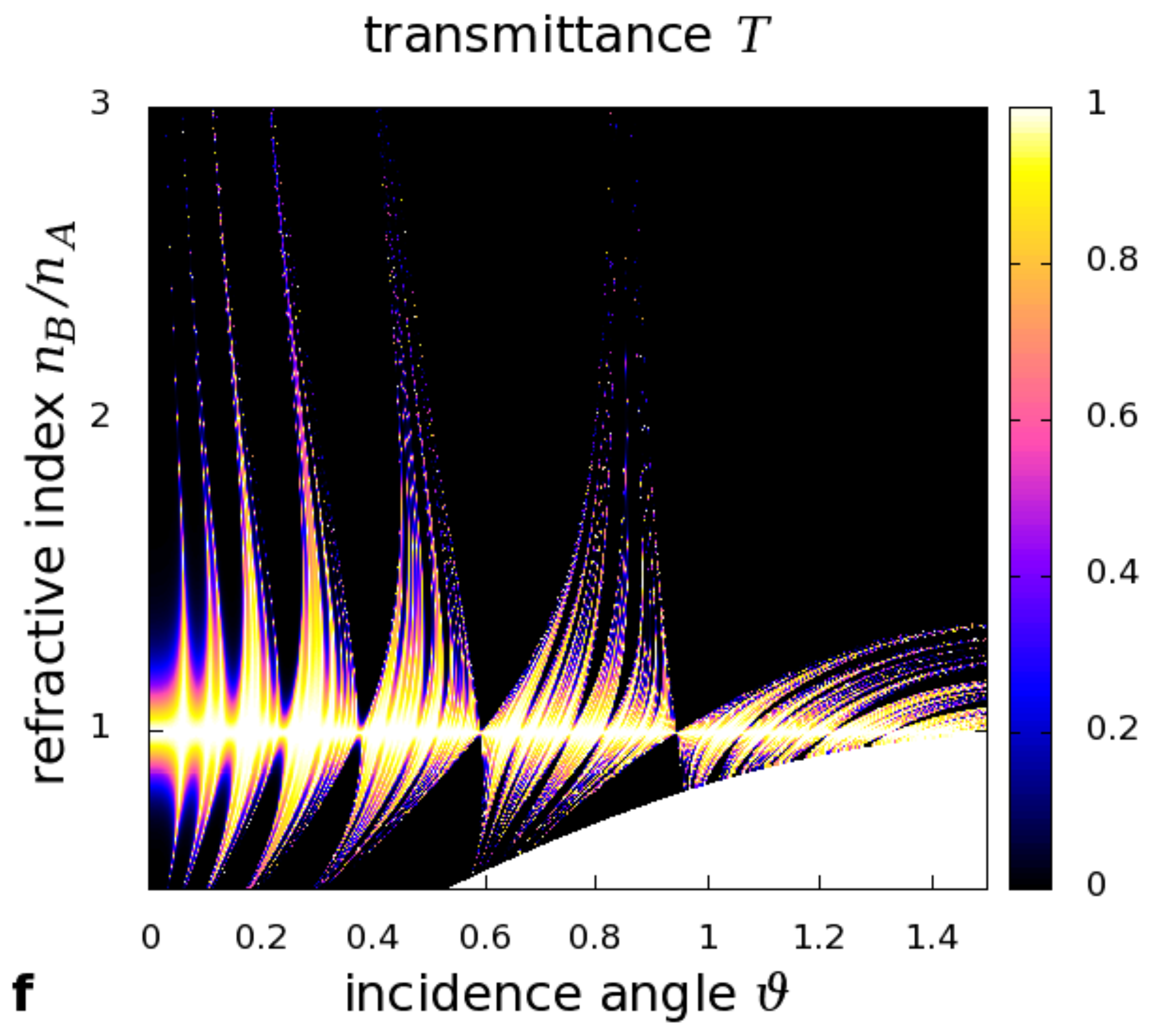}
 \caption{Comparison of the transmittance $T$ of light through a stack made of the materials $A$ and $B$ for the periodic, golden, and silver mean systems, i.e., the same systems as in Figure \ref{fig:bendingT}a, \ref{fig:T_all_angles}a and \ref{fig:T_all_angles}b. Results are shown in panels (a)--(c) in dependence on the reduced frequency $\omega / \omega_0$ and the ratio of the refraction indices $u = n_A / n_B$ for a constant incidence angle $\vartheta = 0.2$ and in panels (d)--(f) in dependency on $u$ and $\vartheta$ for $\omega = \omega_0/2$. The white regions in the lower right corners in (d)--(f) correspond to cases of total reflection.}
 \label{fig:T_all_nratios}
\end{figure*}

In Figure \ref{fig:T_all_nratios} the characteristics of the transmittance $T$ are investigated with respect to the change of the ratio of the indices of refraction $u$. In Figure \ref{fig:T_all_nratios}(a)-(c) we study the influence of the reduced frequency $\omega / \omega_0$ and in Figure \ref{fig:T_all_nratios}(d)-(f) the change of the transmittance $T$ with respect to the incidence angle $\vartheta$ for the periodic, the golden mean and the silver mean sequence.

In the first case, i.e., for the variation of the frequency, it becomes again clearly visible that the quasiperiodicity has the largest effect for $\omega \approx \omega_0/2$. In this region the transmittance $T(\omega, u)$ strongly varies with the used construction rule. By increasing the parameter $a$, more and more bands appear. In the periodic case complete transmission occurs for this frequency only for $u = 1$ , whereas for the quasiperiodic systems there are several bands in this region with a high transmission coefficient. Of course, complete transmission occurs always for $u=1$.

The results for the investigation of the influence of the incidence angle $\vartheta$ for different refraction coefficients are shown in Figure \ref{fig:T_all_nratios}(d)-(f). A comparison of the plots for the periodic and the metallic-mean sequences shows that there is a similar structure of the transmittance $T(\vartheta,u)$ near the regions of total reflection. However, for small incidence angles there is a completely different behavior, i.e., we have zero transmission for $u \ne 1$ in the periodic case, whereas for the quasiperiodic sequences complete transmission occurs for a significant range of ratios $u$.

\section{Conclusion}\label{sec:conclusion}

We investigated the transmission of light through a stack of quasiperiodically aligned materials with respect to the underlying inflation rule and for different incidence angles. One main result for the quasiperiodic systems is that moderately high or even complete transmission can occur for certain ranges of parameters, for which no transmission is found for the periodic system: e.g. we obtained additional bands of moderate transmission for frequencies near the midgap frequency $(b+1/2)\omega_0$ ($b \in \mathbb{N}_0$) with the number of bands increasing depending on the parameter $a$ of the inflation rule, and for a whole range of ratios of refraction indices $u$ even complete transmission occurs.

Further, for all considered systems almost complete transmission occurs for incidence angles $\vartheta$ close to the angle of total reflection. This is caused by the bending of the transmission bands due to the different optical light paths in mediums $A$ and $B$. Additionally, the quasiperiodicity shows only an influence on the transmission coefficients around the region of the midgap frequencies $(b+1/2)\omega_0$, whereas in between the patterns appear to be similar for different quasiperiodic generation rules as well as for the periodic system.

\bibliographystyle{epj}

\begin{thebibliography}{30}

\bibitem{PhysRevLett.1984.Shechtman}
D.~{S}hechtman, I.~{B}lech, D.~{G}ratias, J.W. {C}ahn, {P}hys. {R}ev. {L}ett.
  \textbf{53}(20), 1951 (1984)

\bibitem{PhysRevLett.1987.Wang}
N.~{W}ang, H.~{C}hen, K.H. {K}uo, {P}hys. {R}ev. {L}ett. \textbf{59}(9), 1010
  (1987)

\bibitem{PhysRevLett.1985.Ishimasa}
T.~{I}shimasa, H.U. {N}issen, Y.~{F}ukano, {P}hys. {R}ev. {L}ett.
  \textbf{55}(5), 511 (1985)

\bibitem{JPhysFrance.1989.Sire}
C.~{S}ire, R.~{M}osseri, {J}. {P}hys. {F}rance \textbf{50}(24), 3447 (1989)

\bibitem{PhysRevB.1987.Kohmoto}
M.~{K}ohmoto, B.~{S}utherland, C.~{T}ang, {P}hys. {R}ev. {B} \textbf{35}(3),
  1020 (1987)

\bibitem{JPhysA.1989.Gumbs}
G.~{G}umbs, M.K. {A}li, {J}. {P}hys. {A} \textbf{22}(8), 951 (1989)

\bibitem{JStatPhys.1989.Suto}
A.~{S\"{u}t\H{o}}, {J}. {S}tat. {P}hys. \textbf{56}(3), 525 (1989)

\bibitem{ModPhys.1994.Baake}
M.~{B}aake, U.~{G}rimm, R.J. {B}axter, {I}nt. {J}. {M}od. {P}hys. {B}
  \textbf{8}(25/26), 3579 (1994)

\bibitem{PhysRevLett.2003.Vedmedenko}
E.Y. {V}edmedenko, H.P. {O}epen, J.~{K}irschner, {P}hys. {R}ev. {L}ett.
  \textbf{90}(13), 1372031 (2003)

\bibitem{ZPhysB.1987.Chen}
B.~{C}hen, C.~{G}ong, {Z}. {P}hys. {B}: {C}ondens. {M}atter \textbf{69}(1), 103
  (1987)

\bibitem{PhysRevB.1986.Nori}
F.~{N}ori, J.P. {R}odriguez, {P}hys. {R}ev. {B} \textbf{34}(4), 2207 (1986)

\bibitem{Ferro.2004.Ilan}
R.~{I}lan, E.~{L}iberty, S.~{E}ven-{D}ar {M}andel, R.~{L}ifshitz,
  {F}erroelectrics \textbf{305}, 15 (2004)

\bibitem{PhysWorld.2004.McGrath}
R.~{M}c{G}rath, U.~{G}rimm, R.D. {D}iehl, {P}hys. {W}orld \textbf{17}(12), 23
  (2004)

\bibitem{JPhysD.2007.Steurer}
W.~{S}teurer, D.~{S}utter {W}idmer, {J}. {P}hys. {D}: {A}ppl. {P}hys.
  \textbf{40}(13), R229 (2007)

\bibitem{PhilMag.2008.Bahabad}
A.~{B}ahabad, R.~{L}ifshitz, N.~{V}oloch, A.~{A}rie, {P}hilos. {M}ag.
  \textbf{88}(13--15), 2285 (2008)

\bibitem{PhysRevB.2001.Macia}
E.~{M}aci\'a, {P}hys. {R}ev. {B} \textbf{63}(20), 205421 (2001)

\bibitem{OptExp.2008.Hendrickson}
J.~{H}endrickson, B.C. {R}ichards, J.~{S}weet, G.~{K}hitrova, A.N. {P}oddubny,
  E.L. {I}vchenko, M.~{W}egener, H.M. {G}ibbs, {O}pt. {E}xpress
  \textbf{16}(20), 15382 (2008)

\bibitem{PhysRevLett.1987.Kohmoto}
M.~{K}ohmoto, B.~{S}utherland, K.~{I}guchi, {P}hys. {R}ev. {L}ett.
  \textbf{58}(23), 2436 (1987)

\bibitem{PhysRevE.2009.Esaki}
K.~{E}saki, M.~{S}ato, M.~{K}ohmoto, {P}hys. {R}ev. {E} \textbf{79}(5), 056226
  (2009)

\bibitem{JPhys.2007.Yin}
H.~{Y}in, X.~{Y}ang, Q.~{G}uo, S.~{L}an, {J}ournal of {P}hysics: {C}ondensed
  {M}atter \textbf{19}(35), 356221 (2007)

\bibitem{JPhys.1998.Vasconcelos}
M.S. {V}asconcelos, E.L. {A}lbuquerque, A.M. {M}ariz, {J}. {P}hys.: {C}ond.
  {M}att. \textbf{10}(26), 5839 (1998)

\bibitem{PhysLettA.2004.Li}
J.~{L}i, D.~{Z}hao, Z.~{L}iu, {P}hys. {L}ett. {A} \textbf{332}(5-6), 461 (2004)

\bibitem{JPhys.2006.deMedeiros}
F.F. {de Medeiros}, E.L. {A}lbuquerque, M.S. {V}asconcelos, {J}. {P}hys.:
  {C}ond. {M}att. \textbf{18}(39), 8737 (2006)

\bibitem{PhysRevLett.1994.Gellermann}
W.~{G}ellermann, M.~{K}ohmoto, B.~{S}utherland, P.C. {T}aylor, {P}hys. {R}ev.
  {L}ett. \textbf{72}(5), 633 (1994)

\bibitem{JPhys.2009.Nava}
R.~{N}ava, J.~{Taguena-Martinez}, J.A. del {R}io, G.G. {N}aumis, {J}. {P}hys.:
  {C}ond. {M}att. \textbf{21}(15), 155901 (2009)

\bibitem{PhysRevB.2009.Thiem}
S.~{T}hiem, M.~{S}chreiber, U.~{G}rimm, {P}hys. {R}ev. {B} \textbf{80}(21),
  2142031 (2009)

\bibitem{NonLinAnal.1999.Spinadal}
V.W. de~{S}pinadel, {N}onlinear {A}nalysis \textbf{36}(6), 721 (1999)

\bibitem{Book.Hecht}
E.~{H}echt, A.~{Z}ajac, \emph{{O}ptics}, 7th~edn. (Addison-Wesley, Boston,
  1982)

\bibitem{JPhysA.2006.Honda}
K.~{H}onda, Y.~{O}tobe, {J}. {P}hys. {A} \textbf{39}(20), L315 (2006)

\bibitem{JPhysJap.1991.Kono2}
K.~{K}ono, S.~{N}akada, Y.~{N}arahara, Y.~{O}otuka, {J}. {P}hys. {S}oc. {J}pn.
  \textbf{60}(2), 368 (1991)

\end{thebibliography}

\end{document}